Space-based optical imaging of blue corona discharges on cumulonimbus cloud tops

Yoav Yair[1], Menahem Korzets[1], Adam Devir[2], Melody Korman[3] and Eytan Stibbe[3]

1 – School of Sustainability, Reichman University, Herzliya, Israel

2 – IARD, Kibbutz Yagur, Israel

3 – Rakia Mission, Tel-Aviv, Israel

**\*Corresponding author**
Prof. Yoav Yair
School of Sustainability, Reichman University, (IDC) Herzliya
P.O. Box 167, 8 University Street, Herzliya 4610101 Israel
(p) +972-9-9527952 (m) +972-52-5415091 (f) +972-9-9602401
Email: yoav.yair@runi.ac.il

Keywords: Lightning; Thunderstorms; Transient Luminous Events; International Space Station; Earth Observations

**Abstract.** The ILAN-ES (Imaging of Lightning And Nocturnal Emissions from Space) experiment was conducted in April 2022 as part of the Axiom company AX-1 private mission to the International Space Station, in the framework of Rakia, an Israeli set of experiments selected for flight by the Ramon Foundation and the Israeli Space Agency. The mission objective was to record transient luminous events from the Cupola window in the ISS, based on preliminary thunderstorm forecasts uploaded to the crew 24-36 hours in advance. A Nikon D6 camera with a 50 mm lens was used, in a video mode of 60 fps. During the 12-day mission, 82 different targets were identified for the ISS, of which 20 were imaged by the astronauts, yielding a total harvest > 80 TLEs: sprites, Elves and BLUEs (blue corona discharges). We report here on opportune nadir observation of a thunderstorm that produced multiple blue events near the Myanmar-Thailand border on April 21$^{st}$, 2022, at 21:30 UT. The storm produced many visible blue discharges of varying sizes and durations, in sizes ranging from hundreds of meters to a few km$^2$. The emissions were mostly in blue, however the brightest events had also a conspicuous red component. We used meteorological and ENTLN lightning data to establish the relationship between lightning type and the observable properties of the blue corona discharges.

1. Introduction

Blue optical emissions on the top-most parts of thunderstorms were reported by Lyons et al. (2003) in the framework of the STEPS2000 campaign and nicknamed "pixies". Observed by a ground-based Low-Light TV camera they were described as intense pinpoints of light distributed on the dome of an overshooting cumulonimbus cloud top. There were 83 events in the span of 20 minutes, their size estimated to be 100 m in size. This type of optical emission was different from the 17 upward propagating "gnomes" observed in the same storm, that were probably failed leaders of ordinary lightning discharges propagating upwards (likely the same as the "blue starters" reported by Wescott et al., 1996). In their analysis, Lyons et al. (2003) suggested that these blue discharges may be related to Compact Intracloud Discharges (CIDs), reported by Smith et al. (1999). While there were many ground-based campaigns to observe the various types of transient luminous events in the early 2000s, there were only few reports of pixies and their relationship to lightning, and the microphysical state of the parent thunderstorms when they were generated remained unclear. This may be explained by line-of-sight difficulties and atmospheric absorption of blue emissions.

That observational deficiency was addressed by space-based observations, mostly by the ISUAL satellite, launched in 2004 (Hsu et al., 2017). For example, Liu et al. (2018) used data from the 19 August 2012 storm near lake Taihu in east China and correlated 7 blue events with negative Narrow Bipolar Events (NBEs) observed in VLF radio emissions that occurred within 1 minute. The proximity in time (1 ms) and space (5 km) led Liu et al. (2018) to conclude that negative NBEs are the inception of blue emissions. In a study conducted by Wu et al. (2012) of thousands of NBEs recorded in thunderstorms in China, they were able to establish the existence of two distinct populations of these discharges: positive NBEs that occur, on average, between 8-16 km and mostly negative NBEs occurring higher up, at heights between 16-19 km. That survey confirmed earlier findings reported by Smith et al. (2004) who determined a height of 15 km as differentiating between negative (above) and positive (below) occurrence heights of what was then named "energetic intracloud" (EIC) events. Seemingly, the connection between blue emissions on the tops of thunderstorms and NBEs seems to be well-established, although the physical mechanism was yet to be deciphered.

Similar blue emissions had been observed for the first time from the International Space Station during the IRISS mission in 2015, in the framework of the THOR experiment conducted by Denmark's first astronaut, Andreas Mogensen (Chanrion et al., 2017). In the span of 190 seconds, 245 rapid and short blue emissions were detected visually on the topmost surfaces of two active cumulonimbus clouds. The cells were located over the Bay of Bengal, and based on meteorological data from Chennai, located 200 km southwest of the storm, their tops were at heights between 15.8 -18.2 km. These bluish spots seemed to be appearing at irregular intervals and were located a few kilometers from a strong cloud core of the northern cell, which was higher and reached into the lower stratosphere. The interpretation of Chanrion et al. (2017) to these blue emissions was that they are corona discharges taking place between the uppermost positive charge layer in the cloud close to the anvil, and the screening layer of negative charge accumulating above it.

More recently, the ASIM (Atmosphere-Space Interaction Monitor; Chanrion et al., 2019) payload on the ISS made numerous detections of blue corona discharge events and high-energy emissions (Dimitriadou et al., 2022; Husbjerg et al., 2022; Bjørge-Engeland et al. ,2022; Liu et al. ,2021; Skeie et al., 2022; Neubert et al., 2023), that enabled a thorough characterization of their optical and electrical properties and relations to cloud microphysics and dynamics. Husbjerg et al. (2022) analyzed 3 years of ASIM photometer data (some 57,636 candidates were detected, but the sample was reduced to 11,625 events) and were able to produce the first geographical distribution of blue events, showing maximum occurrence rates in central America (Costa-Rica and Guatemala) and the Caribbean, central and west Africa (Gabon and Cameroon) and the Thailand-Malaysia region in southeast Asia (their Fig. 3). A conspicuous maximum was also observed over the Bay of Bengal (Chanrion et al., 2017) and the Bangladesh-Myanmar coast. They showed that there are two types of blue corona discharges, which differ in their rise-times and presumed location within the thunderstorm: events with fast rise-time of 30 µs or less, that occupy the upper 2 km of the thundercloud top, and those with longer rise-times that are located deeper withing the clouds.

Here we report an optical nadir view observation of a thunderstorm cell that captured - in visible light - the activity of corona blue discharges and lightning, that through the combination of meteorological and lightning location data offer new insights into the relationships between blue events and lightning activity. The superior spatial resolution enabled calculating the accurate size of blue discharges and relate them to cloud features.

2. ILAN-ES experiment design

The ILAN-ES (Imaging of Lightning And Nocturnal Emissions from Space) experiment was conducted in April 2022 by astronaut Eytan Stibbe in the framework of Rakia, an Israeli set of experiments flown on board AX-1, the first private mission to the ISS by Axiom Space. Yair et al. (2023) describe in detail the experiment concept and operational procedures of the astronaut-guided observations towards pre-selected thunderstorm targets, in a similar manner to Chanrion et al. (2017) and Yair et al. (2012). We describe the main components of this experiment here and refer interested readers to those papers. The crew of AX-1 brought a new Nikon D6 to the ISS and it was placed in the Cupola window for earth observations and documentation of the crew's scientific experiments. The camera is slightly superior to the on-board inventory of Nikon D4 and D5 cameras already in the ISS, that show conspicuous degradation of the CCD and many dead pixels. The ILAN-ES configuration during night-time imaging runs was video mode of 60 fps, with a 50 mm lens that had a 26.3°x17.7° FOV (CCD of 1920x1080 pixels). The camera was hand-held and pointed by the astronaut toward the visible lightning, that was within the ~2000km range to the limb (for an average orbital altitude of 410 km). Targets were uploaded 24-26 hours ahead of time and inserted into the Crew Earth Observation (CEO) timeline (see Yair et al., 2023 for details). For nadir view from an ISS orbital height of 420 km, the FOV covers an area of 130x196 km (25,480 $km^2$) and the pixel resolution is 120x102 m (0.01224 $km^{2)}$. Any deviation from a strictly nadir pointing geometry will result in a larger area per pixel, a function of the obliqueness of the viewing angle.

3. Data

In order to enable a meteorological analysis of the conditions prevailing in the areas that were imaged from space, we used the available radiosonde data for the nearest possible station, obtained from the University of Wyoming's Department of Atmospheric Sciences (https://weather.uwyo.edu/upperair/sounding.html). Lightning data was obtained from the Earth Networks Total Lightning Network (ENTLN; Zhu et al., 2021) and the World-Wide Lightning Location Network (WWLLN; Rodger et al., 2006). Satellite imagery was obtained through Eumetsat data services, available at https://view.eumetsat.int/productviewer?v=default and offering various channels in visible and IR, from which cloud top heights could be ascertained. The video was analyzed using the Kinovea video annotation tool, a free software package that enables a frame-by-frame capture and analysis at varying playback speeds (https://www.kinovea.org/). The analysis required manual inspection of the video frames, and in order to avoid subjective biases, was repeated by different researchers to ensure that the identified events were indeed blue discharges and not lightning light with a bluish tinge due to cloud scattering. The detailed analysis of the digital images of blue events was conducted with the ImageJ Pro tool, version 1.54f.

The coordinates of the ISS and calculations of ranges to the prescribed targets were conducted using the Satellite Toolkit software (STK) and verified postflight using the ASIM Science Data Center (ASDC) operated by Denmark Technical University (https://asdc.space.dtu.dk/). This enabled accurate positioning of the ISS orbit with respect to the ground locations of lightning strokes as detected by the lightning networks and correlating the visible lightning in the video with their locations. Notably, since pointing data of the camera was not recorded during observation, it had to be reconstructed based on matching the scenery viewed through the Cupola window with the registration of lightning locations on the ground. The timing of the events, derived from the time stamp of the Nikon D6 camera, was corrected to UTC time based on the procedure described in Yair et al. (2023), where the coincidence of a 124 kA +CG stroke (in ENTLN and WWLLN data) and the induced sprite (in the video) showed a 3.84s difference in the time stamp.

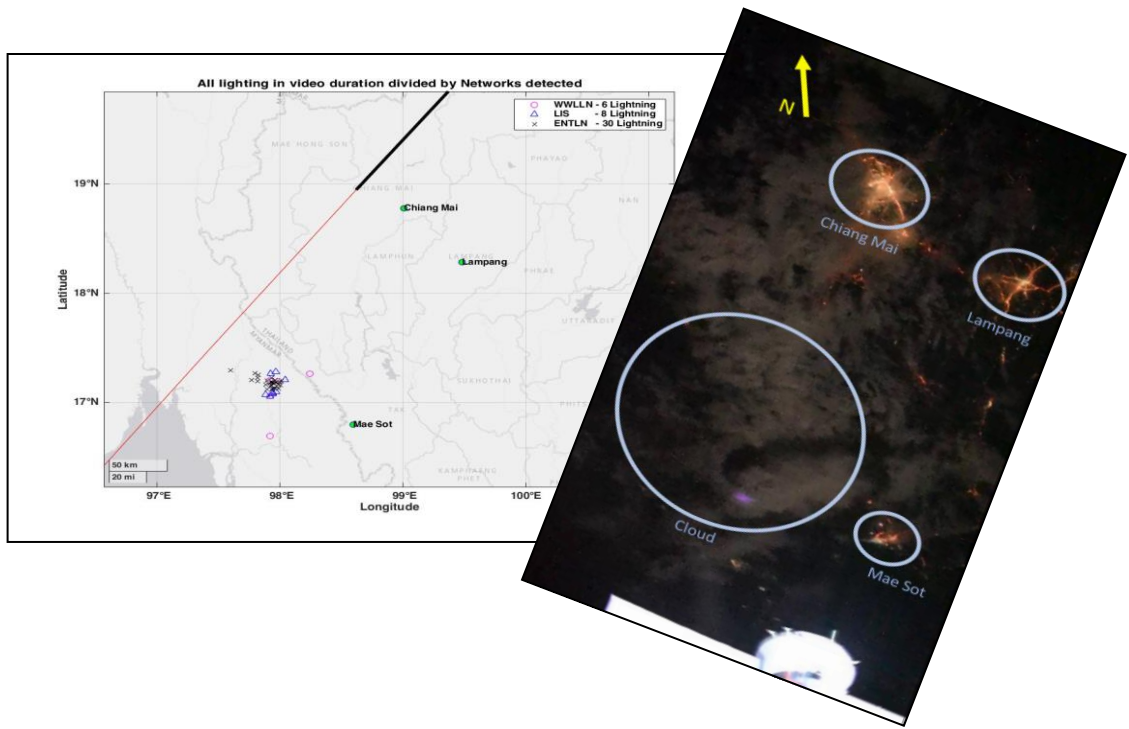

Figure 1: (a) The ISS ground track ascending northeastwards (red and dark diagonal line) on April 21$^{st}$, 2022, 21:30 UT above the Myanmar-Thailand border, with ENTLN, WWLLN and LIS lightning locations. The red segment corresponds to the video timing (b) City lights identified in the video, and the thunderstorm that produced all blue discharges (in blue circle) and a blue event on the upper part.

4. Observations.

4.1 The Myanmar-Thailand border, April 21$^{st}$, 2022, at 21:30 UT

The target was defined as a priority within a list of 10 possible active thunderstorm regions predicted for that day. The initial time for starting the observation was defined 21:17:00 UT near Madagascar, in a northeast direction of the ISS orbit over the Indian Ocean toward the Bay of Bengal and Myanmar, lasting until 21:21:00. Thunderstorms were expected at nadir and port side of the ISS and were imaged in 5 consecutive video segments starting at 21:14:36 north of Madagascar and finishing at 21:35:46 north of the Andaman Sea and the Myanmar-Thailand border, extending into the morning terminator. We focus on the last 30.33 seconds of this orbit, of which only the first 28 seconds were usable (because of sunrise at the ISS). In that short span of time, manual inspection found 18 blue events and 8 lightning flashes. Figure 2 shows the geographical location of the ISS ground track and the lightning detected by the ENTLN, as well as the cities identified in the video. The only available sounding data in the region was from Port Blair (43333 VEPB), India, in the Andaman Sea, approximately 770 km southwest of the observed lightning activity (the station at Chiang Mai (48327 VTCC) which is only 200 km away from the thunderstorm location reported no data for the relevant time). The CAPE was 1679 J kg$^{-1}$, the lifting condensation level is at about 760 hPa, ~2500 m. Above this height, the atmosphere is conditionally unstable all the way up to the equilibrium level at 160 hPa slightly above 14 km, supporting the development of thunderstorms. There was a strong inversion at 18 km, capping further vertical development. Arguably these are not necessarily the conditions prevailing above land in Thailand, but they are still indicative of the regional potential for deep convection. Figure 2 shows a composite IR image of cloud top temperatures. There are two close active thunderstorms, and a zoom-in (Figure 3a) that in the span of 10 minutes between 21:25 and 21:35 UT, produced 265 flashes registered by the ENTLN from these two cumulonimbus cells in the 2°x2° area between 16°N-18°N and 97°E-99°E. At the eastern cell cloud top height was colder than -80°C, higher than 17 km, and based on Figure 2, had a clear indication of overshooting tops at 18.5 km. During that same period of time, the WWLLN detected 54 strokes. When focusing on the specific cell that produced the blue events, ENTLN registered 30 lightning strokes, WWLLN registered 6 and the LIS-ISS detected 8, the same number that appears in the ILAN-ES video.

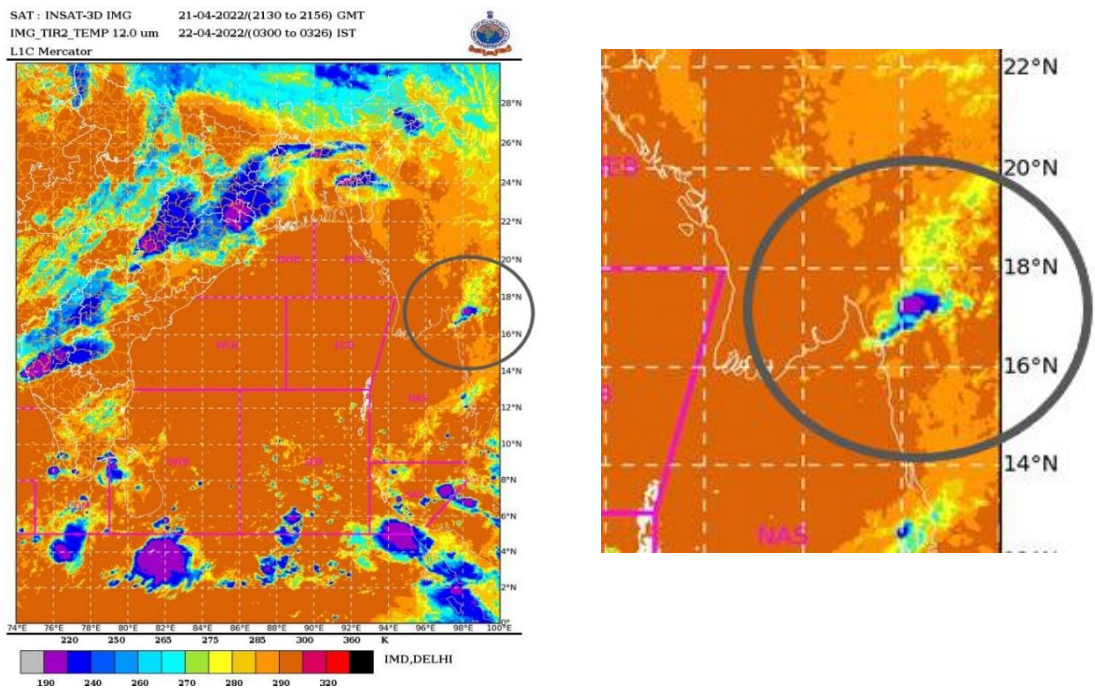

Figure 2: Composite IR 12 µm image by India's INSAT-3D satellite, at 2130 UT on April 21$^{st}$, 2022, with a zoom-in on the cells (right). The active cells on the Myanmar-Thailand border exhibit cloud heights > 14 km (black circle), with overshooting tops colder than 210 K, higher > 18 km.

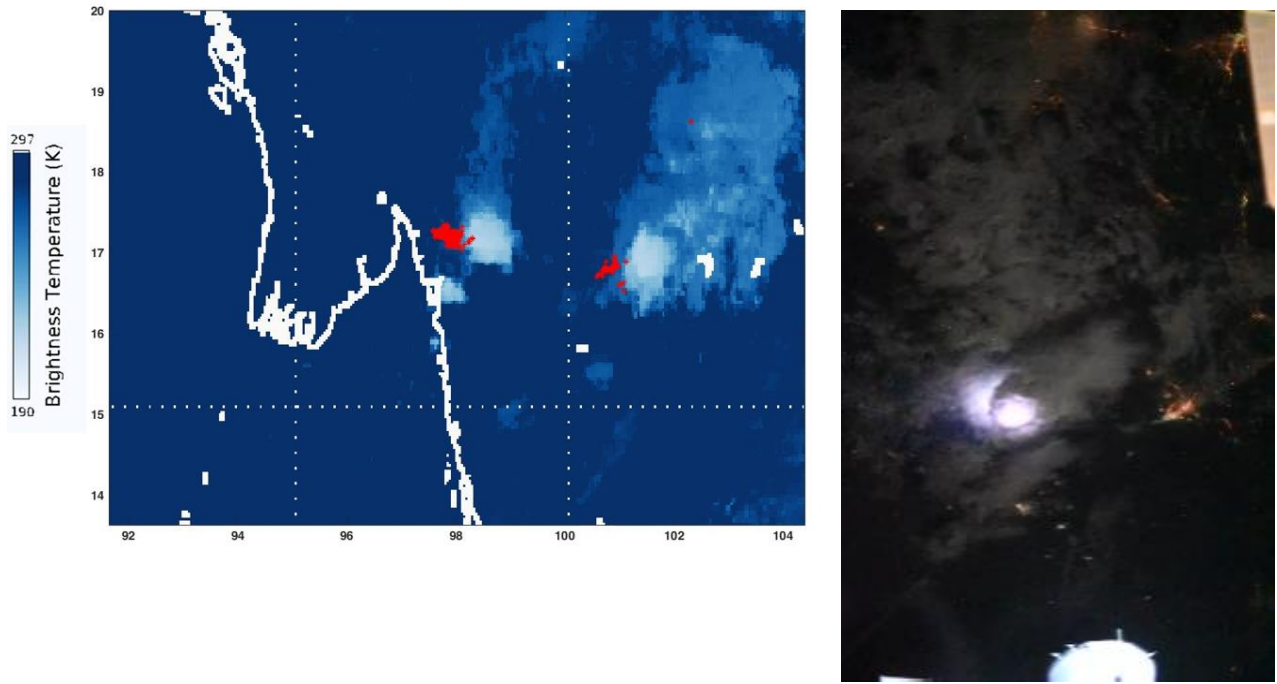

Figure 3: (a) Zoom-in on the two thunderstorm cells near the Myanmar-Thailand border, with superimposed locations of ENTLN strokes during the 10 minute period 21:25-21:35 UTC, April 15th, 2022. (b) Lightning illuminating the overshooting top of the eastern cell, as observed from the Cupola window at 21:30:30.49 UT; Note the ISS solar panel on the top right. Note the Cb turret shadow (black strip north of the flash) cast by the moonlight.

5. Results and Discussion

5.1 Size estimation of BLUE events

The timing of the ILAN-ES campaign in April 2022 coincided with days of significant moonlight, and so the scenery was not fully dark, and clouds were easily discernible even when not illuminated by lightning. The unique nadir observations of the two storm systems enable precise determination of the spatial dimension of the blue corona discharge events, as they appear on top of the cloud. This is done by counting the number of illuminated pixels of the RGB components of the image and subtracting the distribution in an equal area of background pixels of these components. Since the size of each pixel is known, the total count gives the area of the illuminated part. The event in Figure 5 occurred in the Myanmar storm, on April 21st at 22:30 UT, moon phase waning gibbous at 72%. This blue event occurred 2.18 seconds after the beginning of the video, at 21:30:30.02 UT, and lasted 4 frames, a total of 0.0667 seconds. It began as a hardly visible pale blue dot and increased in size and luminosity until reaching maximum size (Fig. 4) that lasted 1 video frame (0.0166 s) and then diminished in brightness and disappeared. The analysis shows that at its peak, the event consists of a central reddish circle, surrounded by a ring of bright blue emission, and a lower arc of reflected blue light from what appears to be a lower laying part of the cloud. This type of event may be defined as a slow blue discharge (Husbjerg et al., 2022). The red/blue pixel number ratio was 0.565, much higher than 0.15 reported by Dimitriadou et al. (2022).

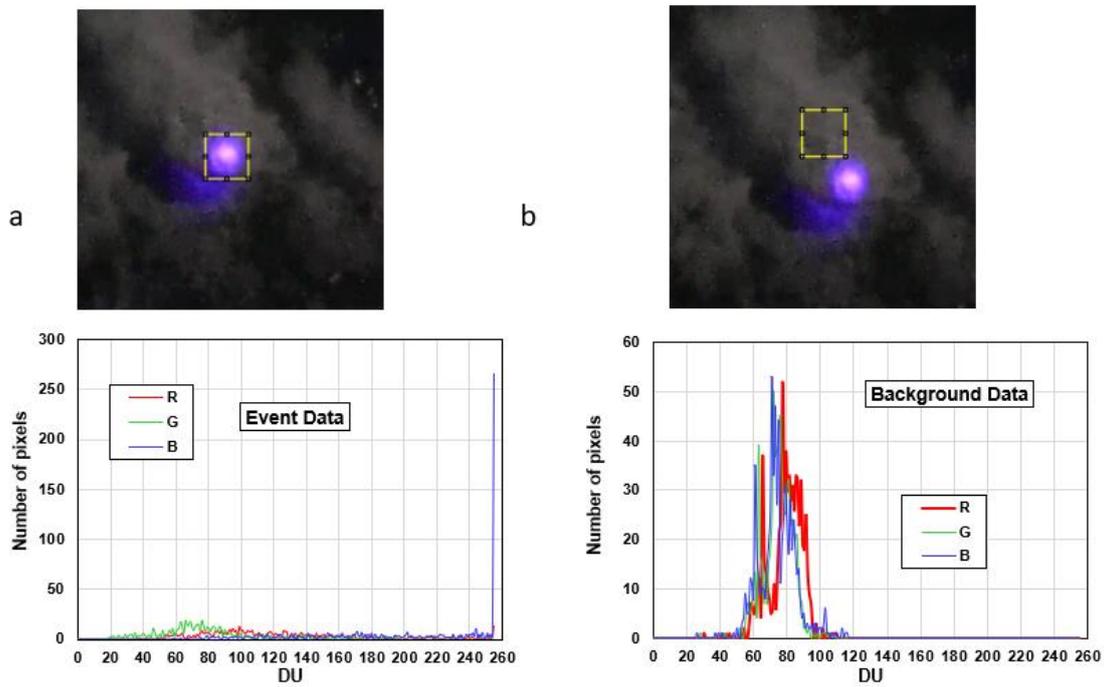

Figure 4: (a) a blue event image and the pixel histogram in the RGB components. After subtracting the background (121 pixels) the red part occupies 368 pixels and the bluish part 653 pixels. (b) The cloud background image and its color histogram, showing nearly equal division of the RGB components.

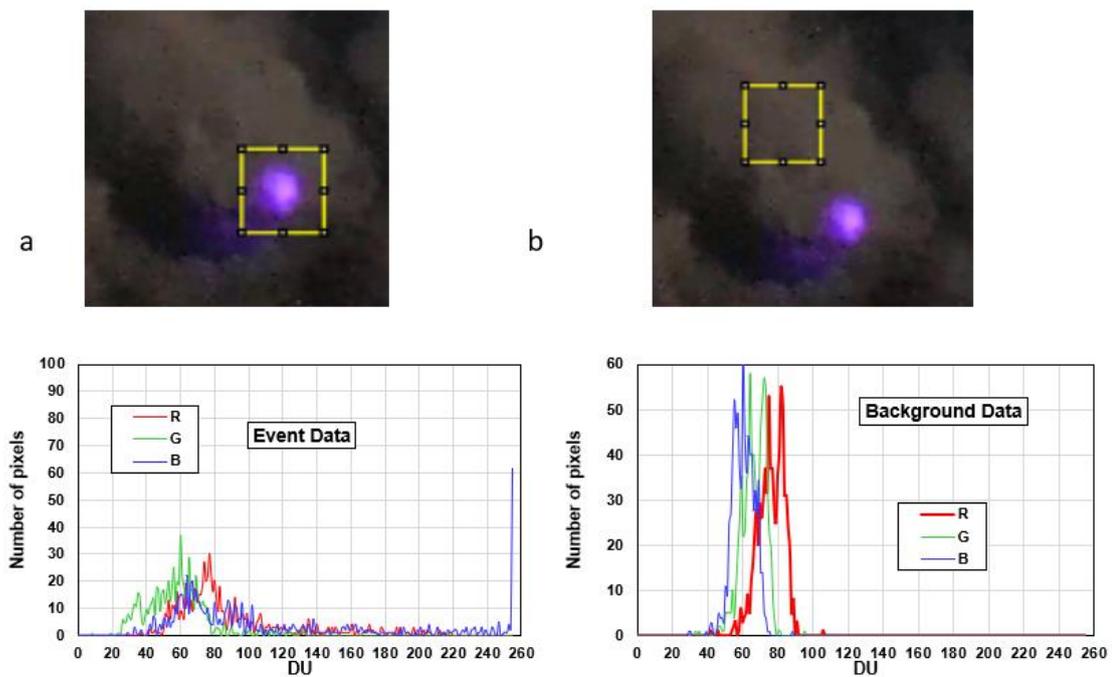

Figure 5: (a) a blue event image and the pixel histogram of the RGB components. After subtracting the background (106 pixels) the red part occupies 172 pixels and the bluish part 276 pixels. (b) The cloud background image and its color histogram, showing nearly equal division of the RGB components.

Based on a size of 0.01224 km$^2$ per pixel, the total illuminated area in the event (blue and red) was 12.49 km$^2$. This is larger than the size deduced by Chanrion et al. (2017) and Dimitriadou et al. (2022), but smaller than reported by Soler et al. (2021), which was 31.4 km$^2$. Another event is shown in Figure 6, found at 14.11 s after the beginning of the video, at 21:30:41.95 UT. This event appeared in only one video frame, 0.0166 s and may be considered a fast event. The total illuminated area by this event was 5.483 km$^2$ and the red/blue pixel number ratio was 0.625.

5.2 Lightning-blue discharge event sequence

The exact relationship between in-cloud discharge process such as Narrow Bipolar Events (either positive or negative), the onset of cloud-to-ground or intracloud discharges and the appearance of blue emissions can be obtained by combining optical (photometric) and lightning detection network data (Soler et al., 2020). Although we lack the superior millisecond time resolution of ASIM (Tables 1 and 2 in Soler et al., 2021; Liu et al., 2022), we were able to create a timeline of the sequence of events identified in the video with 16.66 ms resolution (one frame), and cross correlate the lightning seen in the video with data from other sensors, such as the ENTLN and WWLLN detection systems and the LIS on-board the ISS. In the storm analyzed here, the lightning-BLUEs sequence was complex and there were instances where the discrimination between a diffuse lightning light and a blue corona discharge was hard to perform and can be a subjective definition by the observer. In order to avoid biases, the video analysis was repeated by two independent analysts. In Figure 6 and Table 1 we present the sequence of events as obtained from the ILAN-ES video (after time adjustment, as explained above) for the western cell in the Myanmar-Thailand storm. From the sequence it is evident that blue discharges occur near both cloud-to-ground strokes and intracloud. While lightning typically lasts tens of milliseconds (and thus occupy several video frames), blue discharges are short and frequently appear in a single frame, but some exceptional events appear to last longer. It appears that in most cases, blue events precede lightning by at least 0.0167 seconds (our video resolution). Considering that the average duration of blue events, based on the risetime in the ASIM MMIA data, is ~1 milliseconds (from Figure 4 in Husbjerg et al., 2022), it is beyond the resolution of the video to determine the precise time difference between these two types of discharges.

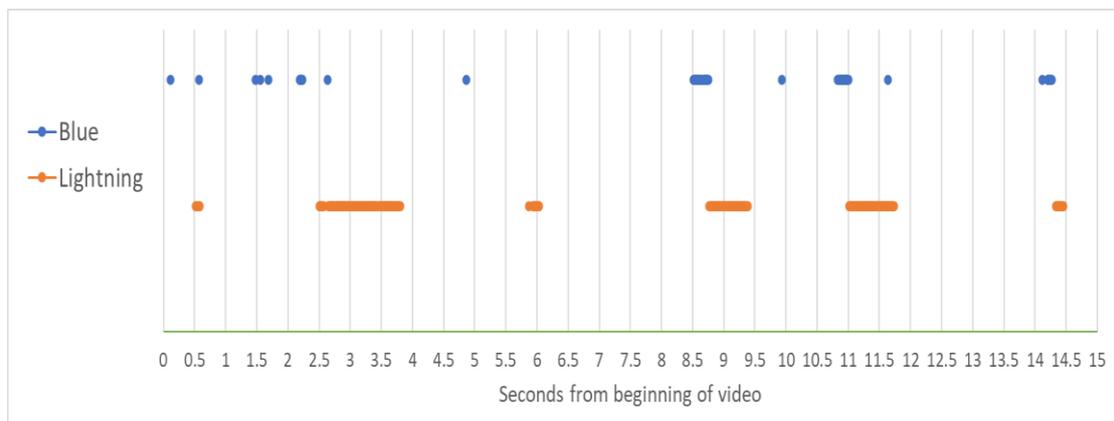

Figure 6: Sequence of events as identified in the video until light from the sunrise entered the camera FOV. The timing of each event depicts the first frame it was discernable in, as detailed in Table 1 (Appendix B).

The first blue event occurs at 21:30:27.96 and was immediately followed by a weak emission from a lightning discharge at 21:30:28.360, that was seen by LIS at 21:30:28.863 and by ENTLN at 21:30:29.138. This was a -10.197 negative IC stroke, with the IC height determined at 11.01 km. There was another blue event on the top of the cloud at the same position, at 21:30:28.410 followed by two frames with white lightning light. After a short hiatus, 3 consecutive blue discharges appeared as small pale blue oval shapes at the outer envelope of the cloud. Then, at 21:30:30.02 appeared a very bright blue event, with an oval shape and a central reddish core. The intensity of this event was strong enough to be reflected from the cloud layer below the turret, creating a ring like blue arc (see Figure 4a). Luminosity lasted 4 video frames (0.067 s). This event was followed 0.32 seconds later by a lightning stroke that lasted 5 frames in the video and appeared as faint white light emanating from the interior of the cloud, likely an IC flash. Another blue discharge followed immediately after (0.14 seconds), preceding a bright cloud-to-ground flash at 21:30:30.490 that illuminated the cloud for 70 video frames (1.167 seconds, until 21:30:31.657, see Figure 3b), with 9 re-brightening events signifying multiple strokes (Quick and Krider, 2013). We suspect that this flash was detected by ENTLN at 21:30:34.419, and was a 12.43 kA negative CG, also observed by LIS at 21:30:34.436 and by the WWLLN at the same time. After a 2-second lull in activity, a small blue dot appeared at 21:30:32.710 preceding a faint lightning stroke, showed as white light below the turret. At 21:30:36.350 an extended blue event appeared as flickering blue oval on the turret, with a total duration of 0.15 seconds, followed by another blue corona discharge at 21:30:36.50, lasting 0.1 seconds. These events are obviously orders of magnitude longer than the typical duration of blue corona discharges, determined to be ~1 ms on average (Soler et al., 2020; Husbjerg et al., 2022).

The reasonable interpretation of these events is that the camera recorded a superposition of multiple blue discharge events occurring in close succession in a confined region of the cloud (Chanrion et al., 2017). A lightning stroke lasting 38 frames (0.633 seconds) appeared immediately after the cessation of blue emissions. This pattern of blue discharges preceding regular lightning appeared again at 21:30:37.78 as blue discharges occurred 180 ms before a bright CG flash (at 21:30:38.85), which lasted 44 frames (0.733 seconds) with 15 re-brightening episodes. It was observed by ENTLN at 21:30:38.853, a +CG with peak current of 10.63 kA. The final sequence of blue discharges – lightning started at 21:30:39.48 with 3 distinct blue events preceded lightning at 21:30:42.17 that lasted for 8 video frames with 3 rebrightening events. That flash may have been the IC stroke detected by ENTLN at 21:30:39.92, that likely occurred deeper in the cloud and was obscured by the large anvil.

6. Summary

Blue emissions at cloud tops are considered to be due to streamers developing in the region between the upper positive charge layer typically located at the anvil of the cumulonimbus and the negative screening layer formed above it in the adjacent air (Krehbiel et al., 2008; Riousset et al., 2010; Liu et al., 2015). They are thought to be related to Narrow Bipolar Events that precede breakdown processes of intracloud strokes in active convective storms (Liu et al., 2018). The overpass of the ISS described here, almost directly above the cumulonimbus cell, occurred during its developing stage, when the convective core produced a clear overshooting top above the anvil, that created an observable shadow on the lower clouds due to the moonlight from above. Subsequent satellite images from 30 minutes after the observation (not shown) exhibit that the area of the anvil expanded and drifted eastward, whereas the cold turret seems to have dissipated and disappeared. This observation confirms the conclusions of Dimitriadou et al. (2022) that showed that blue corona discharges are related to the developing stage of the cumulonimbus in deep convective systems. Furthermore, the windshear the existed in the entire area in the heights of the cloud, and especially above the 150 mb pressure level (14 km) would have induced considerable turbulence at those altitudes, which would support local disruption to the charge structure and promote electrical breakdown (Krehbiel et al., 2008). The results show that blue discharges preceded regular lightning discharges and sometimes were located right above the area where the IC or CG discharge occurred. The advantages of video imaging are clearly exemplified by this example, where even a short 18-second excerpt can help obtain new data. The limited frame rate (60 fps) is better than previous campaigns (Yair et al., 2004, 2011) but lacks the millisecond resolution offered by the new Thor-Davis camera now deployed at the ISS (O. Chanrion, private communication). The success of ILAN-ES further establishes the usefulness of crew-guided observations from space and shows that the Cupola window on ISS is most suitable for lightning and TLE research.

Acknowledgment

The ILAN-ES project was supported by Reichman University Internal Research Fund and by the Canadian Friends of Tel-Aviv University. We wish to thank Eliran Hemo (Ramon Foundation), Brandon Williams (Axiom Space), Jim Watson, Ryan Miller and the TCO team (NASA-MSFC) for their help in mission planning and operation. The student team from the School of Sustainability at Reichman University included Noah Gordis, Anton Teplitskiy, Adam Bernitz, Linoy Levi, Bar Amrami, Bosco Nshimiyumukiza, Yuval Sheleg, Ryan Khalill, Daniel Telkar and Rebecca Mingeleva; Administrative support by Tali Livne.
Shay Katz from the Israeli Meteorological Service helped with the radiosonde data and satellite images, and Martin Stendhal from Denmark's Meteorological Service gave advice on the radiosonde analysis. We thank Timothy Lang from NASA/MSFC for sharing LIS data and to Barry Lynn from Weather-It-Is for the ENTLN data.

List of Figures

Figure 1: (a) The ISS ground track ascending northeastwards (red and dark diagonal line) on April 21st, 2022, 21:30 UT above the Myanmar-Thailand border, with ENTLN, WWLLN and LIS lightning locations. The red segment corresponds to the video timing (b) City lights identified in the video, and the thunderstorm that produced all blue discharges (in blue circle) and a blue event on the upper part.

Figure 2: Composite IR 12 µm image by India's INSAT-3D satellite, at 2130 UT on April 21st, 2022, with a zoom-in on the cells (right). The active cells on the Myanmar-Thailand border exhibit cloud heights > 14 km (black circle), with overshooting tops colder than 210 K, higher > 18 km.

Figure 3: (a) Zoom-in on the two thunderstorm cells near the Myanmar-Thailand border, with superimposed locations of ENTLN strokes during the 10 minute period 21:25-21:35 UTC, April 15th, 2022. (b) Lightning illuminating the overshooting top of the eastern cell, as observed from the Cupola window at 21:30:30.49 UT; Note the ISS solar panel on the top right. Note the Cb turret shadow (black strip north of the flash) cast by the moonlight.

Figure 4: (a) a blue event image and the pixel histogram in the RGB components. After subtracting the background (121 pixels) the red part occupies 368 pixels and the bluish part 653 pixels. (b) The cloud background image and its color histogram, showing nearly equal division of the RGB components.

Figure 5: (a) a blue event image and the pixel histogram of the RGB components. After subtracting the background (106 pixels) the red part occupies 172 pixels and the bluish part 276 pixels. (b) The cloud background image and its color histogram, showing nearly equal division of the RGB components

Figure 6: Sequence of events as identified in the video until light from the sunrise entered the camera FOV. The timing of each event depicts the first frame it was discernable in, as detailed in Table 1 (Appendix B).

Appendix A

Sounding data from Port Blair, India, for April 22$^{nd}$, 2022, 00Z

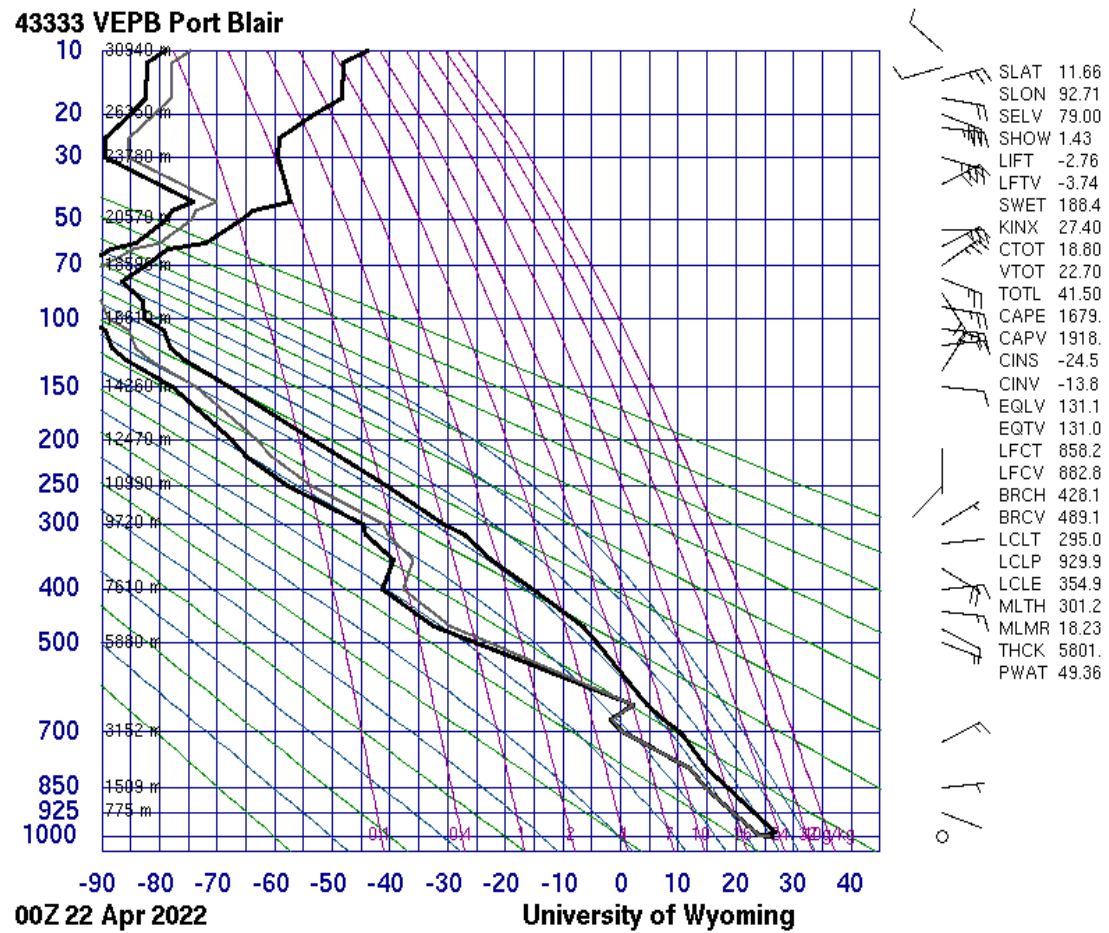

Appendix B

Table 1 – The time sequence and duration of blue events and lightning flashes, for the 18 seconds of the video starting at 21:30:27.84 UT on April 21$^{st}$ 2022.

| # | UTC | Inception Time (s) | End time (s) | Event | Duration (frames) | Duration (seconds) | Description |
|---|---|---|---|---|---|---|---|
| 1 | 21:30:27.840 | | | | | | START Video |
| 2 | 21:30:27.960 | 0.12 | 0.12 | Blue | 1 | 0.017 | Blue appears on cloud top, elongated shape, on turret |
| 3 | 21:30:28.360 | 0.52 | 0.55 | Lightning | 2 | 0.033 | Weak emission from west side of turret |
| 4 | 21:30:28.410 | 0.57 | 0.62 | Blue +lightning | 1+2 | 0.017+0.033 | Bright blue emission circular. Reddish center, followed by white lightning |
| 5 | 21:30:29.320 | 1.48 | 1.48 | Blue | 1 | 0.017 | Small oval shape |
| 6 | 21:30:29.390 | 1.55 | 1.55 | Blue | 1 | 0.017 | Very small dot |
| 7 | 21:30:29.520 | 1.68 | 1.68 | Blue | 1 | 0.017 | Faint, small blue dot |
| 8 | 21:30:30.020 | 2.18 | 2.25 | Blue | 4 | 0.067 | Bright blue, large oval, center red-white, blue crescent below turret |
| 9 | 21:30:30.340 | 2.5 | 2.6 | Lightning | 5 | 0.083 | Intermittent, faint white |
| 10 | 21:30:30.480 | 2.64 | 2.64 | Blue | 1 | 0.017 | Elongated pale blue on top of turret, some light below. |
| 11 | 21:30:30.490 | 2.65 | 3.76 | Lightning | 70 | 1.166 | Continue of [9] Rebrightening at: 2.72; 2.80;2.89;2.92;2.99;3.05; 3.32; 3.35; 3.39; 3.55; 3.62 Multiplicity 9 |
| 12 | 21:30:32.710 | 4.87 | 4.87 | Blue | 1 | 0.017 | Small blue dot on turret |
| 13 | 21:30:33.710 | 5.87 | 5.87 | Lightning | 1 | 0.017 | Very faint white below the turret |
| 14 | 21:30:33.780 | 5.94 | 6.04 | Lightning | 7 | 0.117 | Faint light obscured by cloud, continues from #13 after 3 frames |
| 15 | 21:30:36.350 | 8.51 | 8.64 | Blue | 9 | 0.150 | Flickering blue oval, varying size, top of turret |
| 15 | 21:30:36.350 | 8.51 | 8.64 | Blue | 9 | 0.150 | Flickering blue oval, varying size, top of turret |
| 16 | 21:30:36.500 | 8.66 | 8.74 | Blue | 6 | 0.100 | Same location as #14, flickering, reddish |
| 17 | 21:30:36.600 | 8.76 | 9.38 | Lightning | 38 | 0.633 | Rebrightening at: 8.84; 8.94;8.99; 9.08;9.18;9.24;9.28; Multiplicity 6 |
| 18 | 21:30:37.780 | 9.94 | 9.96 | Blue | 1 | 0.017 | Oval shape, medium size, on top of turret |
| 19 | 21:30:38.670 | 10.83 | 10.99 | Blue | 11 | 0.183 | Pulsating oval on top of turret, Evolving into a flash |
| 20 | 21:30:38.850 | 11.01 | 11.74 | Lightning | 44 | 0.733 | Rebrightening at: 11.04; 11.09;11.14;11.19;11.26;11.29; 11.33; 11.38; 11.41; 11.44; 11.49; 11.54;11.59(max);11.63; 11.68. Multiplicity ~15. |
| 21 | 21:30:39.480 | 11.64 | 11.64 | Blue | 1 | 0.017 | Small blue dot on top turret in the middle of lightning |
| 22 | 21:30:41.950 | 14.11 | 14.13 | Blue | 1 | 0.017 | Concentric, central white on top. Blue light seen below the turret |
| 23 | 21:30:42.040 | 14.2 | 14.25 | Blue | 5 | 0.083 | Elongated blue/red |
| 24 | 21:30:42.170 | 14.33 | 14.43 | Lightning | 8 | 0.133 | First frame appears blue/red. Rebrightening at:14.35; 14.40; 14.43 |
| 25 | 21:30:45.870 | 18.03 | | Sunrise | | | Light coming from northeast as ISS enters dayside |